\documentclass[
	reprint,
	superscriptaddress,
	nofootinbib,
	showkeys,
	aps,
	pra,
	longbibliography
]{revtex4-2}

\usepackage[utf8]{inputenc}

\usepackage[hyperref]{xcolor}
	\definecolor{goethe-blau}{cmyk}{1.0,0.2,0.0,0.4}
	\definecolor{hellgrau}{cmyk}{0.04,0.04,0.05,0.02}
	\definecolor{sandgrau}{cmyk}{0.12,0.09,0.13,0.0}
	\definecolor{dunkelgrau}{cmyk}{0.25,0.25,0.30,0.75}
	\definecolor{purple}{cmyk}{0.08,1.0,0.3,0.36}
	\definecolor{emo-rot}{cmyk}{0.04,1.0,0.8,0.07}
	\definecolor{senfgelb}{cmyk}{0.01,0.25,1.0,0.05}
	\definecolor{gruen}{cmyk}{0.62,0.4,0.87,0.09}
	\definecolor{magenta}{cmyk}{0.08,0.86,0.12,0.12}
	\definecolor{orange}{cmyk}{0.0,0.7,1.0,0.04}
	\definecolor{sonnengelb}{cmyk}{0.0,0.12,0.95,0.0}
	\definecolor{helles-gruen}{cmyk}{0.4,0.17,0.81,0.07}
	\definecolor{lichtblau}{cmyk}{0.8,0.0,0.06,0.04}

\usepackage[
	colorlinks,
	pdfpagelabels,
	breaklinks,
	pdfstartview=FitH,
	bookmarksopen=true,
	bookmarksnumbered=true,
	bookmarksopenlevel=2,
	plainpages=false,
	hypertexnames=false,
	citecolor=emo-rot,
	linkcolor=goethe-blau,
	urlcolor=purple,
	pdftitle={Bound state formation within the Lindblad approach},		
	pdfauthor={Jan Rais, Tim Neidig, Hendrik van Hees, Carsten Greiner},	
	pdfkeywords={}					
]{hyperref}

\usepackage{graphicx}
\usepackage{dcolumn}
\usepackage{braket}
\usepackage[tbtags]{amsmath}
\usepackage{amssymb}
\usepackage{epstopdf}
\usepackage{color}
\usepackage{xspace}
\usepackage[capitalize]{cleveref}
\usepackage{bm}
\usepackage{tabularx}
\usepackage{lipsum}
\usepackage{multirow}
\usepackage[bottom]{footmisc}
\usepackage{orcidlink}
\usepackage{upgreek}
\usepackage{comment}

\usepackage[acronym]{glossaries-extra}
\glssetcategoryattribute{acronym}{nohyperfirst}{true}
\setabbreviationstyle[acronym]{long-short}

\DeclareMathOperator{\Tr}{Tr}
\newacronym{kt}{KT}{Kurganov-Tadmor}
\newacronym{fv}{FV}{Finite Volume}
\newacronym{muscl}{MUSCL}{Monotonic Upstream-centered Scheme for Conservation Laws}
\newacronym{pde}{PDE}{partial differential equation}
\newacronym{cfl}{CFL}{Courant-Friedrichs-Lewy}
\newacronym{clme}{CLME}{Caldeira-Leggett-master equation}

\newacronym{lhs}{l.h.s.}{left hand side}
\newacronym{rhs}{r.h.s.}{right hand side}

\newcommand{\reff}{Ref.~}
\newcommand{\reffs}{Refs.~}

\newcommand{\vdistance}{\vphantom{\bigg(\bigg)}}

\newcommand{\dd}{{\rm d}}

\newcommand{\ee}{{\rm e}}

\newcommand{\kBoltzmann}{k_{\mathrm{B}}}

\newcommand{\REM}[1]{}

\definecolor{magenta}{cmyk}{0,1,0,0}

\allowdisplaybreaks

\begin{document}
	
\title{Lindblad dynamics of deuteron motivated bound states}
	
\author{Jan Rais \orcidlink{0000-0001-8691-6930}}
\email{rais@itp.uni-frankfurt.de} \affiliation{Institut f\"ur
  Theoretische Physik, Johann Wolfgang Goethe-Universit\"at,
  Max-von-Laue-Strasse 1, 60438 Frankfurt am Main, Germany}

\author{Hendrik van Hees \orcidlink{0000-0003-0729-2117}}
\email{hees@itp.uni-frankfurt.de} \affiliation{Institut f\"ur
  Theoretische Physik, Johann Wolfgang Goethe-Universit\"at,
  Max-von-Laue-Strasse 1, 60438 Frankfurt am Main, Germany}%

\author{Carsten Greiner \orcidlink{0000-0001-8933-1321}}
\email{carsten.greiner@itp.uni-frankfurt.de} \affiliation{Institut f\"ur
  Theoretische Physik, Johann Wolfgang Goethe-Universit\"at,
  Max-von-Laue-Strasse 1, 60438 Frankfurt am Main, Germany}


\date{\today}
	
\begin{abstract}
  The Lindblad master equation is a frequently used Markovian approach to describe open quantum systems in terms of the temporal evolution of a reduced density matrix.
  Here, the thermal environment is traced out to obtain an expression to describe the evolution of what is called a system: one particle or a chain of interacting particles, which is/are surrounded by a thermal heat bath.  

  In this work, we investigate the formation of non-relativistic bound states, involving the P\"oschl-Teller potential, in order to discuss the formation time and the thermal equilibrium, applying scales from nuclear physics.
  This problem is borrowed from the field of heavy-ion collisions, where the deuteron is a probe which is measured at temperature regimes around the chemical freeze out temperature, while the deuteron itself has a binding energy which is much lower. 
  This is known and often described as a ``snowball in hell".
  
  We use a reformulated Lindblad equation, in terms of a diffusion-advection equation with sources and therefore provide a hydrodynamical formulation of a dissipative quantum master equation.
\end{abstract}
	

\maketitle

\section{Introduction}
One rather niche area in the large field of open quantum systems is the application of Lindblad dynamics on probes measured in heavy ion collisions.
However, it is not far-fetched to use open quantum systems in this field, because this approach allows to describe a system particle separately from a thermal bath, which is not of further relevance for certain investigations. 
One typical probe, which appears in heavy-ion collisions is the deuteron, a bound state of a proton and a neutron, with a binding energy of approximately 2.3 MeV.
Since, on the one hand, quarkonia, i.e.  bound states of quark-antiquark pairs, are rather successfully described by Lindblad dynamics, cf. \reffs 
\cite{Brambilla:2016wgg,Blaizot:2017ypk,Blaizot:2018oev,Andronic:2024oxz,Brambilla:2022ynh,Brambilla:2017zei,Akamatsu:2020ypb}, a proper theoretical description of the deuteron is under-represented in the literature, which is applying Lindblad dynamics. 
Therefore, let us motivate our main concern: in heavy-ion collisions, energies around the chemical freeze out temperature,  given at $60- 150$ MeV (depending on the collision energies probed in large systems like Au+Au or Pb+Pb) \cite{Braun-Munzinger:2008szb,STAR:2024bpc,Becattini:2009fv}, are reached.  
To measure this matter, colliders as the LHC at CERN or the RHIC at BNL, with collision energies of $\sqrt{s}=2.4-13000$ GeV are used.  
While cooling of the fireball, even at high energies bound states such as heavy quarkonia ($J/\psi$, $\Upsilon$ and excited states) but also the (anti-)deuteron are probes \cite{ParticleDataGroup:2018ovx,Andronic:2017pug}.  
Typical deuteron yields at collision energies of $\sqrt{s_{NN}} = 2.76$ TeV of 
Pb-Pb collision with 0-10\% centrality are $\dd N/\dd y \approx 10^{-1}$
\cite{Andronic:2017pug,ALICE:2015wav}. 
This appears to be remarkable, because the typical binding energy of the deuteron is $\sim 2.3$ MeV, which is orders of magnitude below the hadronic freeze-out temperature. 

For the last decades, several attempts have been made to describe this phenomenon applying various approaches: either by coalescence or by some interference in the interacting matter \cite{Liu:2019nii,Hillmann:2021zgj,Kittiratpattana:2021tpz}.

In this work, we investigate the evolution towards the equilibrium of the bound state and the full system.
We are especially interested, if a Gibbs state is reached in the stationary case, respectively if the density matrix is Boltzmann distributed, and if the time scales, which are typically of the same order of the reciprocal value of the damping are the ones, which are typically observed in heavy-ion collisions. 
Furthermore, we also compare to the equilibration time of the full system, considering the entropy\footnote{Here $S(t)$ is taken to be the standard von-Neumann entropy
	\begin{align}\label{eq:entropy}
		S(t) = \text{Tr}[\rho(t) \ln \rho(t)]\, .
	\end{align}
}, to the equilibration time of certain states.  

We introduce the equations, which we are using to evaluate the Lindblad evolution numerically, and which are known from hydrodynamics.

It turns out, that thermalization is achieved in terms of Boltzmann distributions and that the typical formation time of an arbitrary state\footnote{the minimal value, where $\partial_t \rho_{nn}(t) = 0$} is described by a different time scale, than the full thermalization of the system\footnote{the minimal value, where $\partial_tS(t) = 0$}.

We conclude, that Lindblad evolutions are a useful application to describe bound states.

We work on nuclear scales and set the mass to $m = m_{\text{red,d}} = 470$ MeV, the reduced mass of a deuteron, and $\hbar = \kBoltzmann = 1$.
The size of the computational domain of the  matrix is $L \times L = 40 \text{ fm}\times 40$ fm, which is divided into $300\times 300$ cells to discretize the system for numerical evaluations.

\section{Lindblad equation in a hydrodynamical  formulation}

The general form of the Lindblad-Gorini-Kossakowski-Sudarshan
equation,
\begin{align}\label{eq:Lindblad}
	&\mathcal{L}\left[\rho_S\right] =\\
	&= -\text{i} \left[\tilde{H}, \rho_S\right] +\sum_{i,j=1} \left(L_i \rho_S  L_j -  \frac{1}{2} \left\{L_i^\dagger L_j, \rho_S\right\}\right),\nonumber
\end{align}
contains $i < N$ Lindblad operators $L_i$, where $N$ can be in general the number of possible quantum transitions between states in the system \cite{Koide:2023awf}.
We have implemented a numerical method, which, for the best of our knowledge has not been
used to describe Lindblad dynamics before, but turns out to be a highly efficient tool.
This method has been successfully tested and is discussed in detail in Ref.~\cite{rais2024}.

In \reff \cite{rais2024} we presented a new formulation of the Lindblad equation, \cref{eq:Lindblad}, as an advection-diffusion equation in
conservative form.
Here, the conserved quantity, which has to be satisfied due to the construction of the Lindblad equation is the norm of the density matrix, $\Tr \hat{\rho}=\int \dd x \, \rho(x,x,t)=1$.
We split the density matrix into real and imaginary part, $\vec{u} = \vec{u} ( \vec{x}, t ) = ( \rho_I ( x, y, t ), \rho_R ( x, y,
t ) )^T$, and rearrange the terms of the Lindblad equation in coordinate space representation, which we integrate by parts, cf. \reff  \cite{rais2024}, to obtain
\begin{align}\label{eq:dgl}
	& \partial_t \vec{u} + \partial_x \vec{f}^{\, x} [ \vec{x}, \vec{u} \, ] + \partial_y \vec{f}^{\, y} [ \vec{x}, \vec{u} \, ] =    \vdistance
	\\
	= \, & \partial_x \vec{Q}^{\, x} [ \partial_x \vec{u}, \partial_y \vec{u} \, ] + \partial_y \vec{Q}^{\, y} [ \partial_x \vec{u}, \partial_y \vec{u} \, ] + \vec{S} [ t, \vec{x}, \vec{u} \, ] \, .    \vdistance    \nonumber
\end{align}
Here,  $\vec{f}^{x,y}$, $\vec{Q}^{x,y}$ and $\vec{S}$ are given by

\begin{align}\label{eq:fqs1}
	&\vec{f}^x[\vec{x}, \vec{u}] =\begin{pmatrix}
		- 2 D_{p x} \, ( x - y ) \, \rho_R + \gamma \, ( x - y ) \, \rho_I
		\\
		+ 2 D_{p x} \, ( x - y ) \, \rho_I + \gamma \, ( x - y ) \, \rho_R
	\end{pmatrix}\\
	\label{eq:fqs2}
	&\vec{f}^y[\vec{x}, \vec{u}] =\begin{pmatrix}
		- 2 D_{p x} \, ( x - y ) \, \rho_R - \gamma \, ( x - y ) \, \rho_I
		\\
		+ 2 D_{p x} \, ( x - y ) \, \rho_I - \gamma \, ( x - y ) \, \rho_R
	\end{pmatrix}\\
	\label{eq:fqs3}
	&\vec{Q}^x[\partial_x \vec{u}, \partial_y \vec{u}] =\\
	&=	\begin{pmatrix}
		\frac{\partial}{\partial x} \big[ \frac{1}{2 m} \, \rho_R + D_{x x} \, \rho_I \big] + D_{x x} \, \frac{\partial}{\partial y} \, \rho_I
		\\
		\frac{\partial}{\partial x} \, \big[ - \frac{1}{2 m} \, \rho_I + D_{x x} \, \rho_R \big] + D_{x x} \, \frac{\partial}{\partial y} \, \rho_R
	\end{pmatrix}\nonumber\\
	\label{eq:fqs4}
	&\vec{Q}^y[\partial_x \vec{u}, \partial_y \vec{u}]= \\
	&=	\begin{pmatrix}
		\frac{\partial}{\partial y} \big[ - \frac{1}{2 m} \, \rho_R + D_{x x} \, \rho_I \big] + D_{x x} \, \frac{\partial}{\partial x} \, \rho_I
		\\
		\frac{\partial}{\partial y} \, \big[ \frac{1}{2 m} \,  \rho_I + D_{x x} \, \rho_R \big] + D_{x x} \, \frac{\partial}{\partial x} \, \rho_R
	\end{pmatrix}\nonumber\\
	\label{eq:fqs5}
	&\vec{S}[t, \vec{x},\vec{u}] =\\
	&=	\begin{pmatrix}
		( V ( y ) - V ( x ) ) \, \rho_R + \big[ 2 \gamma - D_{p p} \, ( x - y )^2 \big] \,\rho_I
		\\
		( V ( x ) - V ( y ) ) \, \rho_I + \big[ 2 \gamma - D_{p p} \, ( x - y )^2 \big]\rho_R
	\end{pmatrix}\, ,\nonumber
\end{align}
where $m$ is the mass of the system particle, $V(x)$ the systems potential, $\gamma$ the damping coefficient, satisfying the dissipation-fluctuation theorem, and therefore is related to the friction $\eta$ \cite{Caldeira:1982iu}, and $D_{pp}, D_{px}$ and $D_{xx}$ the diffusion coefficients, which are detailed in \reffs \cite{BRE02,DEKKER19811,Homa2019,Bernad2018} and generally depend on  the heat bath temperature $T$, and $\Omega$, the cutoff frequency of the Ohmic bath spectrum, as well as $m$ and $\gamma$.
To numerically solve \cref{eq:dgl}, we use a finite-volume scheme, the Kurganov-Tadmor scheme,  which is introduced and explicitly discussed in \reff \cite{KURGANOV2000241}.
For further details, we refer to \reff \cite{rais2024}.


\section{Bound state formation and formation time}

		
\begin{figure*}
	\begin{center}
		\hspace*{\fill}%
		\includegraphics[width=\linewidth]{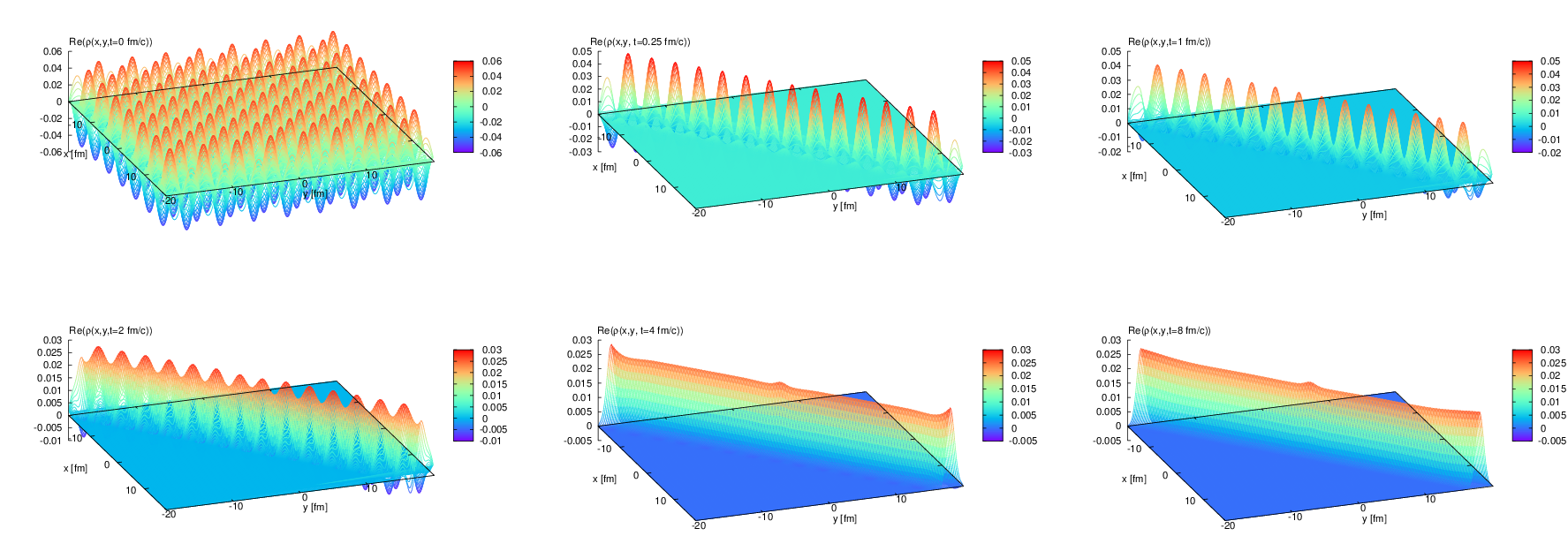}
		\hspace*{\fill}%
		
		\caption{
			The temporal evolution of $\rho(x,y,t)$, with initially populated $16^\text{th}$ state towards thermal equilibrium at times $t=0$ fm/c, $t=0.25$ fm/c, $t=1$ fm/c, $t=2$ fm/c, $t=4$ fm/c and $t=8$ fm/c of the real part of the reduced density matrix. 
			The bath temperature is $T=250$ MeV, $\Omega = 4T$, $\gamma = 0.1$ MeV and $D_{px} = - \gamma T/ \Omega$, which $V(x)$, the potential given in \cref{eq:potential}. 
		}
		\label{fig:3d_init10}
	\end{center}
\end{figure*}

\begin{figure}
	\begin{center}
		\hspace*{\fill}%
		\includegraphics[width=\linewidth]{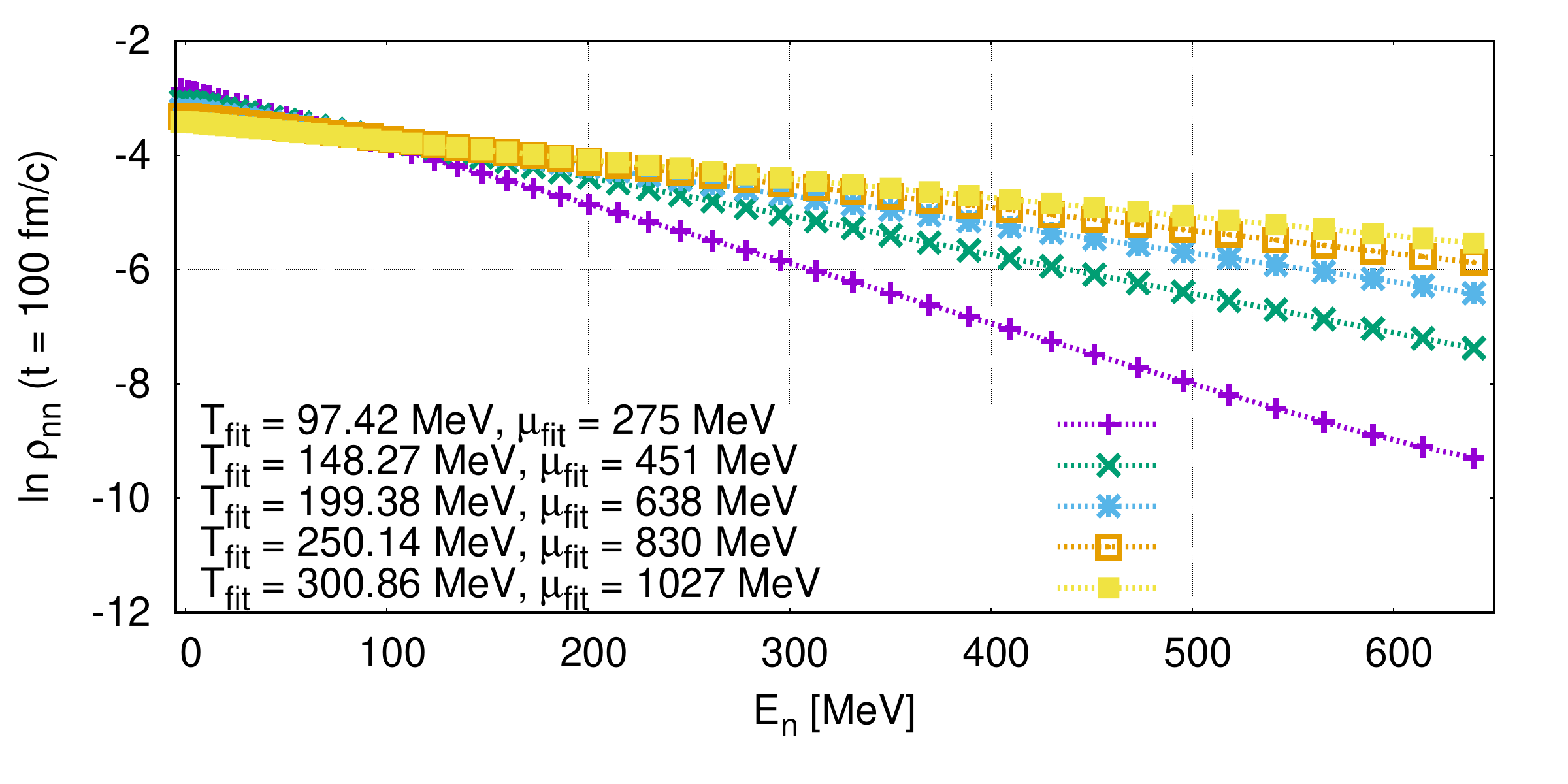}
		\hspace*{\fill}%
		\caption{
			Logarithm of the final distribution of $\rho_{nn}(t)$ at time $t_{\text{eq}} = t = 100$ fm/c for different heat bath temperatures $T = 100,150,200,250,300$ MeV. 
			Here, $\gamma = 0.1$ c/fm, $\Omega = 4T$, $D_{px} = 0$ and the initial condition is $n=8$. The dashed line illustrate the fit curves obtained by applying \cref{eq:Boltzmann}. 
		}
		\label{fig:Boltzmann_fit}
	\end{center}
\end{figure}
\begin{figure}
	\begin{center}
		\hspace*{\fill}%
		\includegraphics[width=1.0\columnwidth,clip=true]{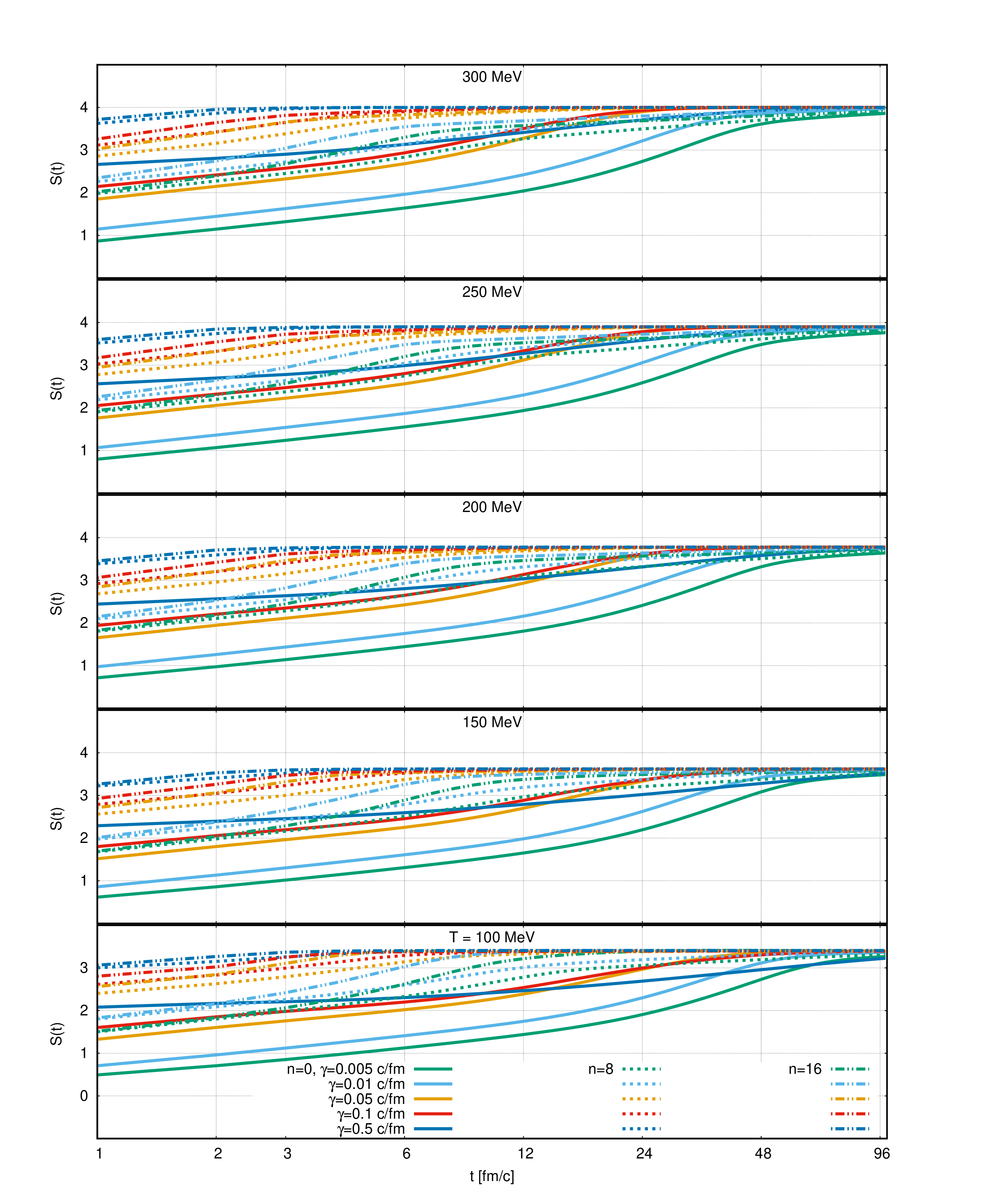}
		\hspace*{\fill}%
		
		\caption{
			Entropy $S(t)$ for different bath temperatures $T$ for the pure Caldeira-Leggett master equation, where $D_{px}=0$ and $\Omega = 4 T$. The different line colours correspond to different damping $\gamma$, while the different line-styles correspond to different initial conditions $n=0,8,16$.
		}
		\label{fig:entropy_4T_0Dpx}
	\end{center}
\end{figure}

\begin{figure}
	\begin{center}
		\hspace*{\fill}%
		\includegraphics[width=1.0\columnwidth,clip=true]{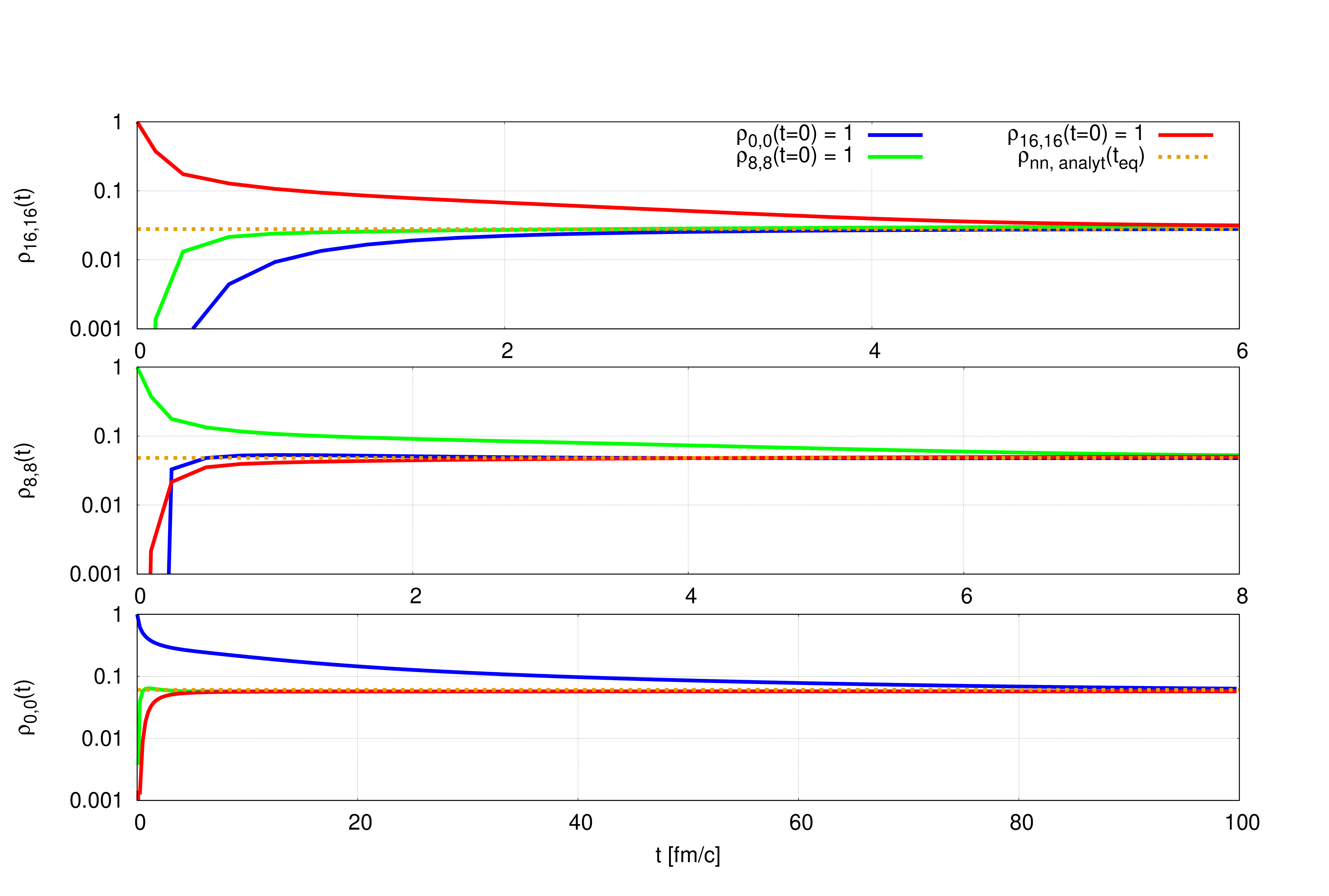}
		\hspace*{\fill}%
		
		\caption{
			$\rho_{0\,0}(t)$, $\rho_{8\, 8}(t)$ and $\rho_{16\,16}(t)$ depending on different initial conditions, $\rho_{nn} = 1$ with $n=0,8$ and $16$ for bath temperatures $T = 100$ MeV for the pure Caldeira-Leggett master equation, where $D_{px}=0$ and $\Omega = 4 T$ and damping $\gamma=0.1$ c/fm. The dashed line illustrates the equilibrium result, which corresponds to the fit of the Boltzmann distribution, \cref{eq:Boltzmann}, as indicated in \cref{fig:Boltzmann_fit}.
		}
		\label{fig:gournd_state}
	\end{center}
\end{figure}

In this chapter we discuss the bound-state problem in order to provide a description of  the formation of non-relativistic particles in heavy-ion collisions as it has been motivated in \reff  \cite{Rais:2022gfg}. We are using the potential
\begin{align}\label{eq:potential}
	V(x) = \begin{cases}
		-V_0\,\frac{1}{\cosh^2(\alpha x)}, &\qquad\text{if } \vert x\vert \leq 20 \text{fm}\\
		\infty& \qquad\text{if } \vert x\vert  > 20 \text{fm}\, ,\\
	\end{cases}
\end{align}
which is introduced in \reff  \cite{rais2025boundstate}.
Here, $\alpha = 1/r_\text{d}$, where $r_\text{d}$ is the deuteron radius.
We use the energy eigenfunctions, which are calculated via a shooting method numerically, to define the initial conditions by
\begin{align}
	\rho(x,y,0) = \sum_{m,n=0}^N c_{mn} \braket{x\vert \psi_m}\braket{ \psi_n\vert y}\, ,
\end{align}
where $N$ is the highest considered state, and therefore the state, where the Hilbert space of the system particle is truncated (here $N=50$, such that the energy spectrum is $E\in \left[-2.3 \text{ MeV},  640\text{ MeV}\right]$). 
In \reff  \cite{rais2025boundstate} we provide an extensive analysis of the dependency of the system on the parameters $T$, $\gamma$, $\Omega$, $D_{px}$ and  the initial condition. 
In \cref{fig:3d_init10} we illustratively show the temporal evolution of the real part of the reduced density matrix, where, as a initial condition, the $16^\text{th}$ state (which has an energy of approximately 60 MeV) is populated.
  
We use 
\begin{align}\label{eq:coeff}
	\rho_{nm}(t) = \int\dd x \int \dd y \, \rho(x,y,t) \braket{\psi_n\vert x}\braket{y \vert \psi_m}\,
\end{align}
to calculate $\rho_{mn}(t)$, which allows to investigate the temporal behavior of each state, the decoherence of the system, and the final distribution, which is compared to
\begin{align}\label{eq:Boltzmann}
	\rho_{\text{analyt.}, nn} = \exp\left[-\frac{1}{T} \left(E-\mu\right)\right]\,.
\end{align}

\subsection{Thermalization of the bound state problem}\label{sec:thermalization}

For late times of the simulation, where the equilibration is expected, we can calculate $\rho_{nn}(t)$ to compare the result to \cref{eq:Boltzmann}, which is used as a fit function with fit parameters $T_{\text{fit}}$, which is expected to coincide with the heat bath temperature, and $\mu_{\text{fit}}$, the chemical potential.
In \reff \cite{rais2025boundstate} we compare various results for different values of $\gamma, T, \Omega$, and different values of $D_{px}$, which we also compare to the pure Caldeira-Leggett master equation, where $D_{px}=0$. For mathematical reasons, cf. \reff \cite{rais2024}, we set $D_{xx} = 0$.
One result for the equilibrium distributions of $\rho_{nn}(t=t_{\text{eq}})$ for various heat bath temperatures is depicted in \cref{fig:Boltzmann_fit}.

One can see a nice agreement between the heat bath temperature $T$, and the fitted temperature $T_{\text{fit}}$ which shows, taking into account, that also the distribution of $\rho(x,-x,t)\sim \ee^{-2mTx^2}$, cf. \cite{rais2025boundstate} is fulfilled, and that the system equilibrates with the heat bath, as expected.

\subsection{Entropy}\label{sec:entropy}

Diagonalizing the density matrix allows us to calculate the entropy, \cref{eq:entropy}, and to study the dynamics towards equilibrium of the full system. 
This is depicted in \cref{fig:entropy_4T_0Dpx} for various initial conditions, heat bath temperatures and dampings. 
In \cref{fig:gournd_state} we show the dynamics of some explicitly taken states, which we also use to set up the initial conditions. 
One can see, that contra-intuitively the states do not decay strictly exponentially but several time scales are involved during the evolution. 
Furthermore, one can see, that the thermalization takes place faster for higher states, and the bound state, especially, if it is initially fully populated, takes longest to relax.
Therefore, the time scales of each state are different and dependent on the initial condition. 
Also the full equilibration of the system, which can be seen in \cref{fig:entropy_4T_0Dpx}, shows a different time scale.

The dashed line in \cref{fig:gournd_state} provides a comparison to the expected equilibrium distribution, calculated using the fits from \cref{eq:Boltzmann}.

\section{Summary}
In this work, we summarized the most important findings of \reffs \cite{rais2024,rais2025boundstate}, where we have discussed the formation or destruction time of a bound state with parameters mimicking a deuteron; incorporating nuclear scales and energies.
To this end, we used Lindblad dynamics and rewrote the Lindblad equation into a conservative advection-diffusion equation to solve the temporal evolution numerically.
We found, that equilibration with the heat bath is reached.
 Therefore, the equilibration time is dependent on the heat bath temperature, the damping, as well as the cutoff frequency of the Ohmic bath spectrum. 
Futhermore, we found, that the pure Caldeira-Leggett master equation leads to a thermal state as well as the Lindblad-type structures, that are used frequently, and the results in the most cases do not differ significantly, cf. \cite{rais2025boundstate}.

Even though the damping dictates the thermalization time, it is also sensitive to the initial condition.
The thermalization time of the full system and therefore the minimal time, where an entropy maximum is reached is smaller, if an energetically higher state is originally populated, because the gap between the energy of the system and the environment is smaller. 
Furthermore, the thermalization time of an arbitrarily considered state is higher, if the energy gap between this state and the initial condition is small. 

To conclude -- it is possible to stably form a bound state  due to environmental effects, with a probability, which is sensitive to the heat bath temperature, the damping and the initial energy in the system.
Therefore, Lindblad dynamics provide a quantum mechanical  explanation of the ``deuteron paradox".

\begin{acknowledgements}
	
	J.~R.\  acknowledges support by the \textit{Helmholtz Graduate School for Hadron and Ion Research for the Facility for Antiproton and Ion Research} (HGS-HIRe for FAIR) and the \textit{Deutsche Forschungsgemeinschaft} (DFG, German Research Foundation).
	
	J.~R.\ thanks J.P.~Blaizot for having the opportunity of in person discussions, creative ideas and collaboration.
	
	J.~R.\ thanks N.~Zorbach and  A.~Koenigstein for developing a highly reliable numerical code, which was used to perform all simulations. Forthermore,  J.~R.\ thanks for the professional collaboration.
	
	J.~R.\ thanks T. Neidig  for creative and useful discussions and impulses.
\end{acknowledgements}

\appendix

\bibliography{bibliography.bib}

\end{document}